\documentclass[review]{elsarticle}
\usepackage{amsmath}
\usepackage{graphicx}
\usepackage{hyperref}

\journal{Comptes rendus de l'Académie des Sciences}

\begin{document}

\begin{frontmatter}

\title{Towards hybrid circuit quantum electrodynamics with quantum dots}

\author{J.J. Viennot$^{1}$, M.R. Delbecq$^{2}$, L.E. Bruhat, M.C. Dartiailh, M. Desjardins, M. Baillergeau, A. Cottet and T. Kontos\fnref{myfootnote}}
\address{Laboratoire Pierre Aigrain, Ecole Normale
Sup\'{e}rieure-PSL Research University, CNRS, Universit\'{e} Pierre et Marie
Curie-Sorbonne Universit\'{e}s, Universit\'{e} Paris Diderot-Sorbonne Paris
Cit\'{e}, 24 rue Lhomond, 75231 Paris Cedex 05, France}
\fntext[a]{now at JILA, University of Colorado, Boulder, Colorado 80309, USA}
\fntext[b]{now at RIKEN, Japan}
\fntext[myfootnote]{to whom correspondence should be addressed : kontos@lpa.ens.fr}

\begin{abstract}
Cavity quantum electrodynamics allows one to study the interaction between light and matter at the most elementary level. The methods developed in this field have taught us how to probe and manipulate individual quantum systems like atoms and superconducting quantum bits with an exquisite accuracy. There is now a strong effort to extend further these methods to other quantum systems, and in particular hybrid quantum dot circuits. This could turn out to be instrumental for a noninvasive study of quantum dot circuits and a realization of scalable spin quantum bit architectures. It could also provide an interesting platform for quantum simulation of simple fermion-boson condensed matter systems. In this short review, we discuss the experimental state of the art for hybrid circuit quantum electrodynamics with quantum dots, and we present  a simple theoretical modeling of experiments.\end{abstract}


\end{frontmatter}


\section{Interest of the use of cQED techniques}

Using an oscillator as a detector to readout the state of a system coupled to it is a widely used method in classical and quantum physics. Some of the most renowned example include Atomic Force Microscopy [1], Nuclear Magnetic Resonance spectroscopy [2] or mass sensing. In circuit quantum electrodynamics (\emph{cQED}), these resonators can be in the quantum regime with discrete photon states interacting with mesoscopic circuits. For devices such as quantum dots, the high frequency capacitive measurement offered by superconducting resonators is to be compared to the usual electronic transport techniques. In contrast with conductance measurements, it provides a fast and non invasive detection and allows measurements of devices with extremely small couplings to leads, thus avoiding decoherence caused by coupling to fermionic reservoirs. This issue was previously overcome using techniques such as radio-frequency charge sensing [3]. However, high finesse, high frequency ($GHz$) resonators go beyond and naturally provide a high sensitivity and high speed measurements. Working at high frequencies also allows one to measure effects related to quantum capacitance [4,5], or investigate electronic transitions resonant with the cavity. By avoiding transport, cQED readout techniques could yield to QND [6,7] (quantum non-demolition) measurements of charge or spin states in quantum dot devices. Another significant potential of cQED architectures is the scalability which, combined with recently demonstrated spin-photon coupling [8], could allow us to tackle fundamental problems such as the coupling and entanglement of distant spins [9,10,11,12,13,14,15,16]. Finally, because modern nano-fabrication techniques, in general, allow one to build devices with arbitrary complexity, cQED provides a conciliating platform to go towards hybrid systems which combine different types of quantum degrees of freedom and their respective advantages. Superconducting circuits have already been used to couple microwave photons to large spin ensembles [17,18], magnons [19], but also mechanical resonators [20,21] or optical photons [22].

	\section{Technical realization of hybrid circuit quantum electrodynamics devices with quantum dots}
	
The realization of circuit quantum electrodynamics architectures with quantum dot circuits has been enabled by the progress of nanofabrication techniques. This combines low dimensional conductors with metallic electrodes and superconducting resonators. Quantum dot circuits based on GaAs 2-dimensional electron gases [23], Semi-conducting Nanowires [24], carbon nanotubes [25], and graphene [26] have already been coupled to cavities. Depending on host materials, fabrication methods vary and present different challenges when it comes to obtaining high quality factors. For instance GaAs 2-dimensional electron gases must be be kept away from the resonator field to avoid dissipation.  GaAs substrates also have piezoelectric properties that can cause microwave loss. Carbon nanotubes require a step of chemical vapor deposition growth at high temperature under a hydrogen atmosphere, and the growth is associated with deposition of amorphous carbon that can cause strong dissipation. Various techniques have thus been developed to circumvent these issues [23,25,26].

	\section{Coupling to individual electronic states in quantum dots} \label{Section:CouplingIndividualStates}

\begin{figure}\centering
\includegraphics[width=0.75\textwidth]{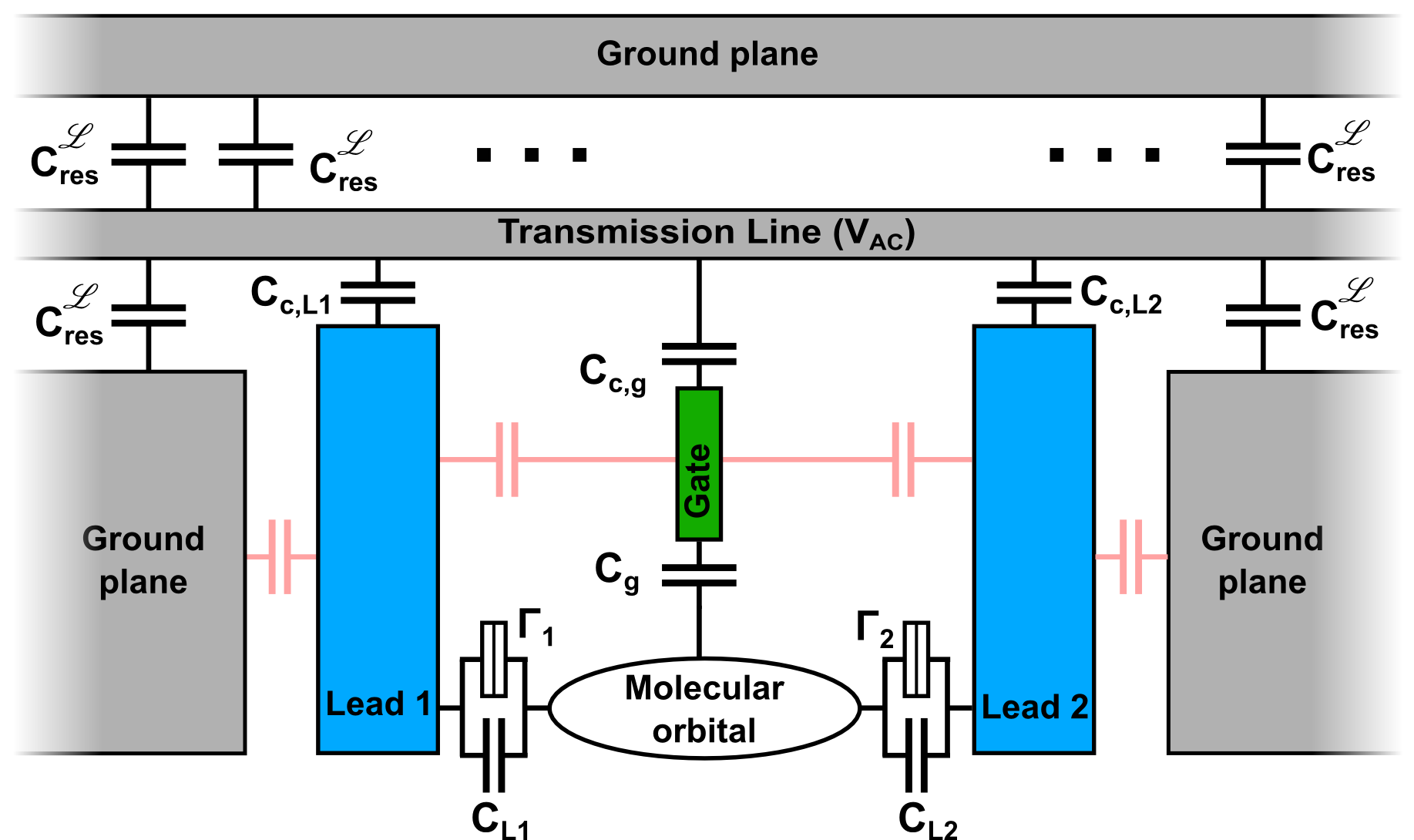}
\caption{Scheme of a dot system hosting an electronic orbital and coupled to the transmission line of a co-planar wave-guide resonator.  The different capacitances possibly involved are summarized, plus some potential parasitic capacitances (light pink). $C^\mathcal{L}_{res}$ is the linear capacitance of the resonator to ground $\left( \int C^\mathcal{L}_{res} = C_{res} \right)$. The capacitances of the electrodes to the transmission line can be replaced by galvanic connections to increase coupling strength and/or selectivity. Connections to DC voltage sources and associated capacitances are not represented. }
\label{Figure:DQDRES:CouplingCapaSingle}
\end{figure}

Quantum dot individual states are naturally electrically coupled to the
electromagnetic field via their charge density. In a first approach, one may
use a circuit diagram as in figure \ref{Figure:DQDRES:CouplingCapaSingle} to
explain how the quantum dot interacts with the field trapped in a transmission
line resonator. The quantum dot capacitance $C_{dot}$ is typically of the
order of the $aF$. The total inductance $L_{res}$ of the transmission line and
its total capacitance $C_{res}$ to the ground are typically $L_{res}%
\approx0.5nH$ and $C_{res}\approx1pF$ (see e.g. [27]). A small change in
$C_{res}$ due to the quantum dot total capacitance leads to a change in
$f_{c}$ by $\Delta f_{c}$, i.e.%

\begin{equation}
f_{c}+\Delta f_{c} = \frac{1}{2\pi\sqrt{L_{res}(C_{res}+\Delta C_{dot})} }
\approx f_{c} \left(  1-\frac{\Delta C_{dot}}{2C_{res}} \right)
\end{equation}

\begin{equation}
\Delta f_{c}\approx-\frac{f_{c}}{2C_{res}}\Delta C_{dot}\label{EquationDeltaC}%
\end{equation}
It is always practical to think in terms of an effective dot capacitance in
order to evaluate the orders of magnitudes involved in a given experiment.
However, one should keep in mind that the picture of Fig.1 is a strong
approximation since quantum dots circuits are in general non-linear systems
which can for instance be used to obtain a lasing effect (see section 6.4).

Microscopically, the coupling between a simple quantum dot circuit (with no
loops) and a cavity can be described by incorporating an electric potential
term in the cavity+dot hamiltonian [28]:
\begin{equation}
H_{dot-cavity}=e\int d^{3}r\hat{\rho}(r)v(r)V_{rms}(a+a^{\dagger
})\label{couplingdot+cav}%
\end{equation}
where $\hat{\rho}(r)$ is the electronic charge density operator on the quantum
dot, $v(r)$ is a form factor accounting for the mode geometry, $V_{rms}%
(a+a^{\dagger})$ is the usual quantized form of the cavity central conductor
potential, and $e$ the elementary charge. In Eq. \ref{couplingdot+cav}, we
restrict ourselves to the case of a single mode for simplicity. As shown in
ref [28], Eq. \ref{couplingdot+cav} can lead to photo-induced tunneling terms
which go beyond a capacitive circuit model. However, in simple situations
where the tunnel coupling between the different circuit elements and/or the
reservoirs density of states are sufficiently low, these terms can be
disregarded. In this limit, Eq. \ref{couplingdot+cav} leads to a picture
qualitatively similar to the scheme of Fig.1, where the potential of each
electronic orbital is shifted linearly by the cavity potential. Two important
cases then derive from the coupling hamiltonian \ref{couplingdot+cav}: the
single dot case where
\begin{equation}
H_{dot-cavity}=eV_{rms}\alpha_{d}\hat{n}_{d}(a+a^{\dag})\label{H1dot}%
\end{equation}
and the double dot case where
\begin{equation}
H_{dot-cavity}=eV_{rms}\left(  \alpha_{L}\hat{n}_{L}+\alpha_{R}\hat{n}%
_{R}\right)  (a+a^{\dag})\label{HcDQD}%
\end{equation}
with $\hat{n}_{d/L/R}$ the operators associated to the total number of
electrons in the dots. Above, the coefficients $\alpha_{d}$, $\alpha_{L}$ and
$\alpha_{R}$ are assumed to be orbital-independent in each quantum dot, for
simplicity. The coupling terms \ref{H1dot} and \ref{HcDQD} lead to Eq.
\ref{EquationDeltaC} in simple situations, like for instance if the cavity
frequency is the smallest scale in the problem [29]. The term $\Delta C_{dot}$
has generally two contributions, one arising from the geometrical capacitances
connected to the dots and one ``quantum capacitance'' contribution due to the
finite density of states in the dots. However, if the cavity frequency is
larger than the tunnel rate between the dot and the reservoirs, one may
instead find an inductive dot contribution (see section 4.2).

In practice, the cavity frequency shift can be determined by measuring how the
cavity reflects or transmits a microwave signal. Both the phase and the
amplitude of the microwave output signal can be measured. In this review, we
will mainly discuss the behavior of the phase signal, which directly reveals
the cavity frequency shift. The amplitude of the output microwave signal can
reveal cavity damping induced by the quantum dot circuit. A complete
characterization of the photon statistics can also be performed with the
cavity output field tomography, which corresponds to a quasi-probabilistic map
of the two output field quadratures (see Figure \ref{Figure:Liu}).

    \section{Coupling of single quantum dots}	
    
High frequency resonators provide a powerful tool to investigate the physics of quantum dots coupled to fermionic reservoirs. The presence of these coupled reservoirs can affect the way electrons in the dot couple to photons, but it can also deeply modify the behavior or electrons in the dot. It can for instance lead to the emergence of many body effects such as the Kondo effect, of which the high frequency dynamics can be studied using a resonator [30]. Importantly, the ratio between the resonator frequency and the dot-lead coupling rate is a determinant factor for the system dynamics [31]. This problem has been recently revisited in the case of a single dot coupled with normal and superconducting contacts [32]
The regime in which the dot-lead coupling rate is dominant could be instrumental for the study of charge relaxation in an open quantum dot [28]. It is expected to be universal at low temperature in the single contact limit [4,5], considering the coupling scheme of section \ref{Section:CouplingIndividualStates}.

	\subsection{Coupling to open quantum systems}
	
\begin{figure}\centering
\includegraphics[width=0.55\textwidth]{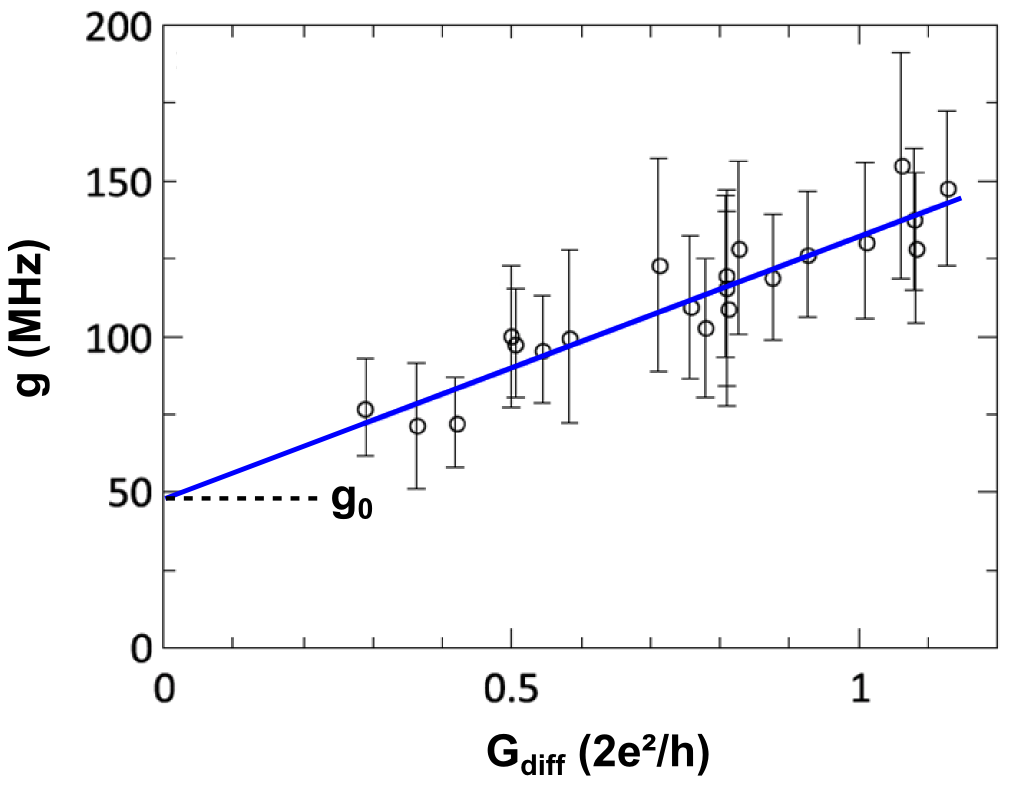}
\caption{Effective dot-resonator coupling strength of a single quantum dot as a function of the dot's differential conductance. For this specific device, coupling strength increases with conductance, therefore with $\Gamma_{Lead}$. The intercept at zero conductance (closed dot) gives an estimate of the geometric bare coupling strength $g_0$. Source: [33]}
\label{Figure:DQDRES:CouplingGamma}
\end{figure}

A quantum dot coupled to leads (\emph{i.e. fermionic reservoirs}) is an open quantum system with a finite quantum capacitance [4,5]. In the case of large coupling to the leads $\Gamma_{lead} > f_c$, this contributes to the total capacitance of the dot and therefore renormalizes the coupling to the resonator. Figure \ref{Figure:DQDRES:CouplingGamma} shows an example of such a behavior for the effective dot-resonator coupling $g$ as a function of lead coupling rate $\Gamma_{Lead}$ (considering that $G_{diff} \propto \Gamma_{Lead}$). This measurement was performed on a device where the coupling to the resonator was dominated by capacitances to the leads. In this case, the increase of $\Gamma_{Lead}$ leads to an increase of the effective lever-arm of the lead over the dot, hence the increase of the effective dot-resonator coupling.

	\subsection{Coupling to a nearly closed quantum dot}
	
\begin{figure}\centering
\includegraphics[width=0.55\textwidth]{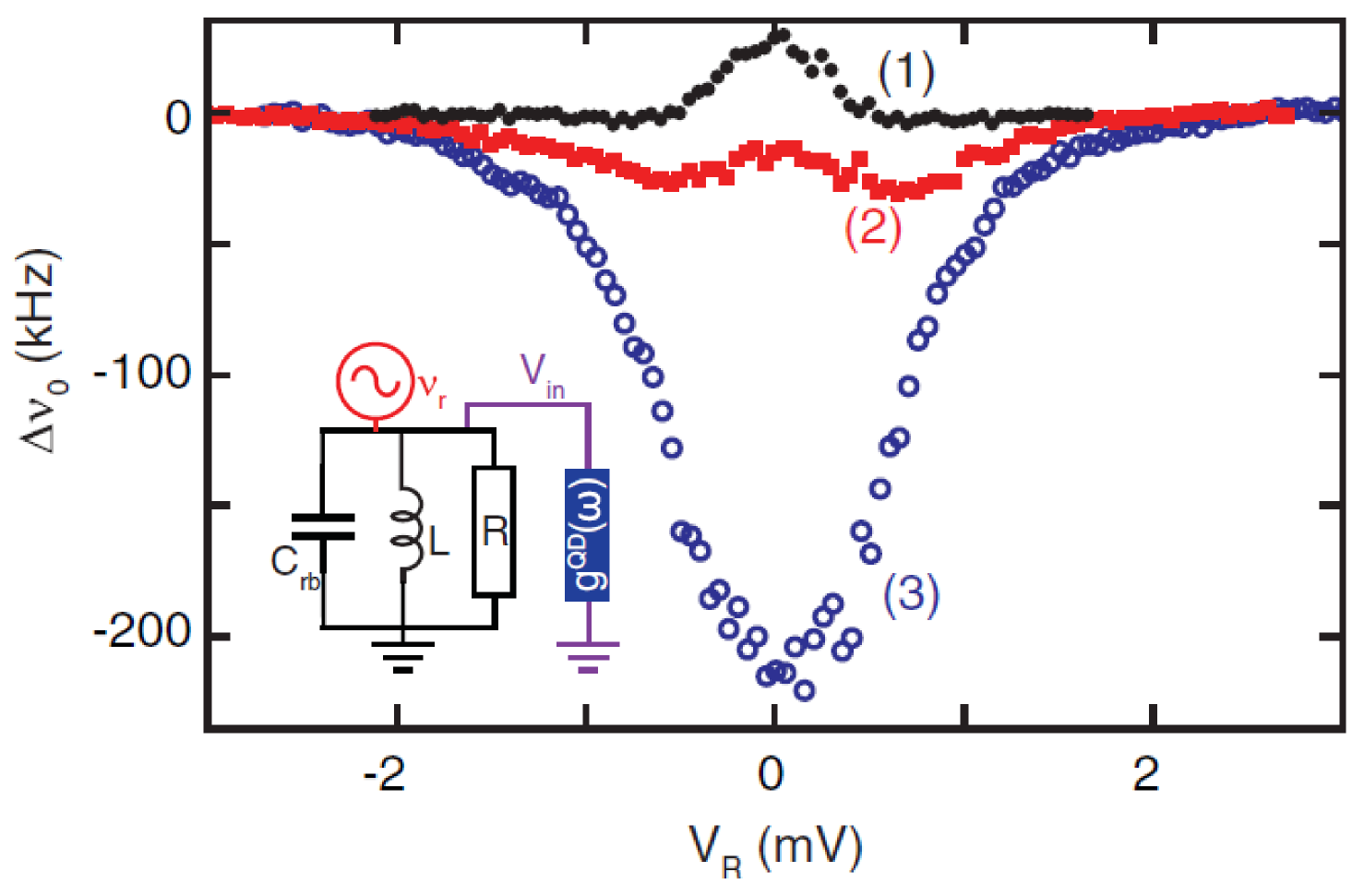}
\caption{Resonator frequency shift (here referred to as $\Delta \nu_0$) as a function of the effective single dot gate voltage $V_R$, for three different coupling rates $\Gamma_{lead}$. Estimated $\Gamma_{lead}$'s are (1)$20MHz$, (2)$58MHz$ and (3)$125MHz$. $f_c=6.7GHz$. A negative $\Delta \nu_0$ corresponds to a dot capacitive behavior and a positive $\Delta \nu_0$ to a dominantly inductive behavior. Inset: equivalent circuit with quantum dot admittance $g^{QD}(\omega)$ connected to the resonator lumped element circuit. Source: [31]}
\label{Figure:DQDRES:Frey2012}
\end{figure}

In the opposite case where coupling to the fermionic leads is small, $\Gamma_{lead} < f_c$, one can look at the resonator response when dot-lead transitions take place. This has been studied with a GaAs 2DEG-based effective single dot - single lead device gate-coupled to a resonator [31]. As shown in figure \ref{Figure:DQDRES:Frey2012}, the sign of the cavity frequency shift $\Delta f_c$ changes with the characteristic dot frequency $\Gamma_{lead}$. In this experiment, data is interpreted in a non interacting formalism using scattering matrix theory [4], recalling the model of the quantum capacitance. As $\Gamma_{lead}$ decreases and becomes comparable to the resonator frequency, the response of the dot system presents a crossover from capacitive to inductive. While the capacitive response is interpreted as the usual quantum capacitance contribution due to the low density of states in the quantum dot, the inductive component is interpreted as a lagging effect occurring because the dwell time of electrons on the dot is larger than $f_c^{-1}$.

	\subsection{Coupling distant quantum dots}

Although the interaction between the resonator photons and a single dot dot-lead transition is in the weak coupling regime, it can be enough to mediate an interaction between distant quantum dots. Similarly to condensed matter situations, where several strongly correlated electronic orbitals can be coupled to bosonic modes (such as phonons), two artificial orbitals are coupled to a photonic mode [33]. The resulting interaction has the form of a polaronic shift. The orbital energy of a dot is thus shifted by a quantity proportional to the number of electrons in the distant dot and the coupling strengths of the resonator to the two dots. This is a first step towards quantum simulations, where the quantum properties of the device under study can be engineered in order to emulate a more complex quantum mechanics problem.

	\section{"Dipolar" coupling of double quantum dots}
		\subsection{Devices and coupling schemes}
	
\begin{figure}\centering
\includegraphics[width=1.0\textwidth]{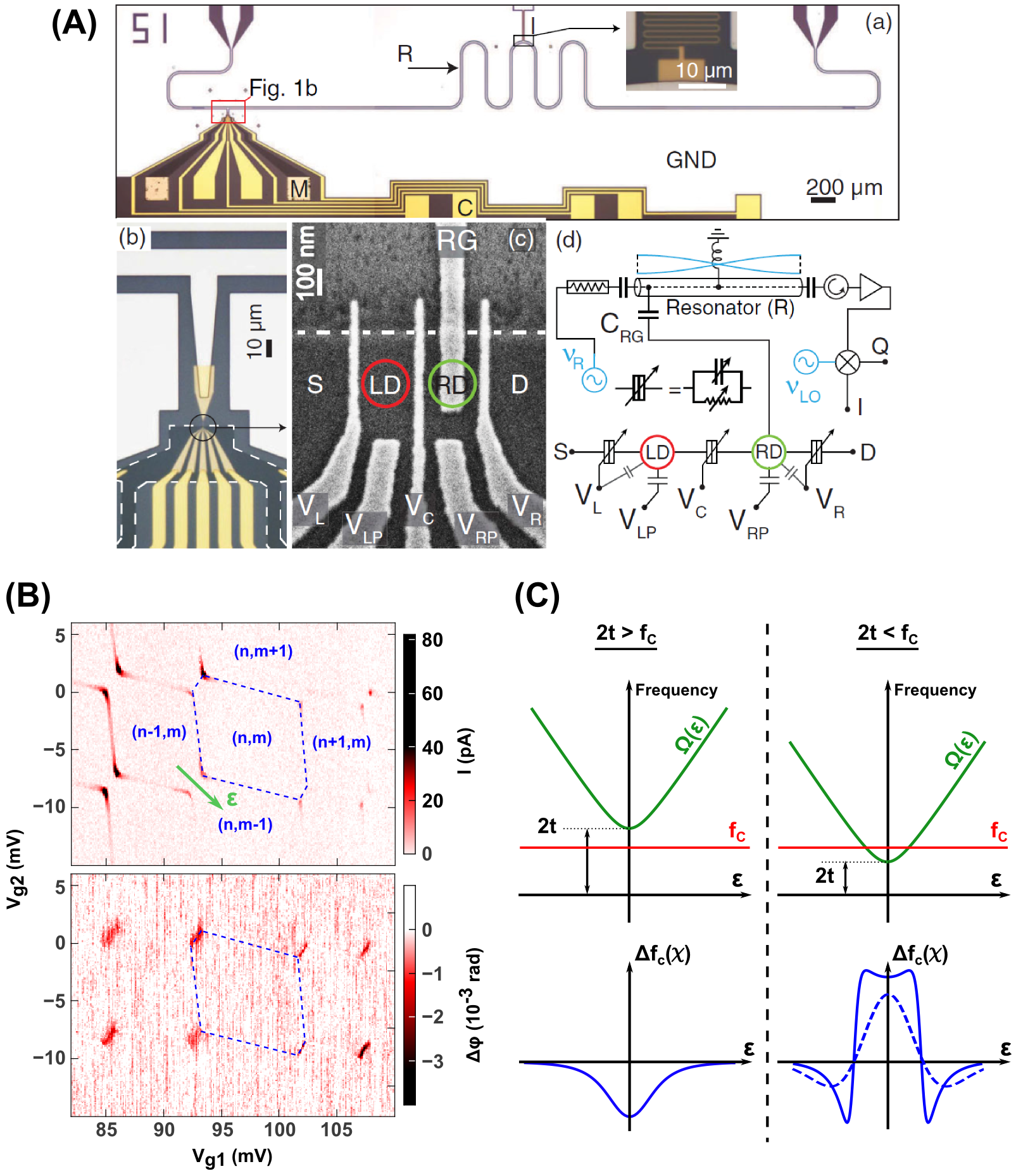}
\caption{\textbf{(A)} Coupling geometry for a GaAs double quantum dot and a superconducting Al resonator. Source: [36]   \textbf{(B)} Stability diagram of a double quantum dot obtained via DC current and resonator transmission phase measurements. Source: [25] \textbf{(C)} Dispersion relation $\Omega(\epsilon)$ and corresponding charge susceptibility $\chi$ of a DQD charge qubit, readout through a coupled cavity, in the linear regime. When $2t<f_c$, cavity and qubit spectral lines cross in the weak coupling regime ($g_0 < \Gamma_2$). The susceptibility changes sign with detuning $\Delta$ and qualitatively depends on whether decoherence is strong (dashed line) or weaker (full line).}
\label{Figure:DQD}
\end{figure}

Compared to single quantum dots, double quantum dots (DQD) have the crucial advantage to have gate tunable internal (inter-dot) electronic transitions. As one changes the gate voltages of the two dots, electrons tunnel out or in the dots and fill up their orbitals. The charge stability diagram of a DQD thus exhibits regions where transitions between different charge states are possible [34]. The stability diagram of a device can be obtain by sweeping local gate voltages while measuring the low frequency conductance of the device, or measuring the transmission of the coupled resonator (see figure \ref{Figure:DQD}B). Inter-dot transitions happen near the so-called triple points, along the zero detuning lines, where orbital of the two dots hybridize. These transitions can therefore be tuned in or out of resonance with the resonator frequency. Coupling to these transitions requires a very asymmetric capacitive coupling to each of the dots (for instance $\left| \alpha_L- \alpha_R \right| \simeq \alpha_L $), in order to couple and modulate the inter-dot energy detuning. This can be achieved in different ways and an example of coupling geometry is given in figure \ref{Figure:DQD}A. Figure \ref{Figure:DQD}.A.a shows a large scale optical micrograph of the resonator. The double dot is located in the area surrounded by the red rectangle, enlarged in Figs.\ref{Figure:DQD}.A.b and c. The gate RG visible in Fig \ref{Figure:DQD}.A.c has been designed to enhance the coupling between the right quantum dot RD and the cavity. This gate is galvanically coupled to the resonator central conductor, as visible in Fig. \ref{Figure:DQD}.A.b. 

Close to degeneracy between the two dots (small inter-dot detuning $\epsilon$), the orbitals of the dots hybridize into bonding and anti-bonding states [35]. Although it has a relatively short coherence time, this charge doublet can be viewed as a qubit that is naturally coupled to the resonator [24,36,37,38]. The spectrum of this qubit and the resulting resonator response for qubit in the ground state ($\langle \sigma_z \rangle =-1$) are given in figure \ref{Figure:DQD}C. In these experiments coupling to the lead $\Gamma_{Lead}$ is made as small as possible to reduce relaxation to the reservoirs, and the coupling strength only depends on capacitance ratios via geometrical facts.

		\subsection{Equation of motion for a DQD charge qubit in a cavity} \label{SectionChargeEOM}

We write here the equation of motion of the operators involved in the system Hamiltonian to calculate the cavity frequency shift in both the dispersive and resonant regimes. The DQD states form a charge qubit which can be readout using the cavity dispersive shift, but also driven out of equilibrium by DC transport or cavity photons. This treatment, introduced in Ref. [37] thus goes further than the theory developed for instance in [24,36,39] by integrating the non-linear regime with $\langle \sigma_z \rangle \neq -1$.

Let us start with the Hamiltonian of the system with the ingredients necessary to conveniently describe electronic transport and out of equilibrium effects. The relevant states of the DQD are: $\{|\emptyset\rangle,|B\rangle ,|AB\rangle , |2\rangle \}$ where $|\emptyset\rangle$ and $|2\rangle$ are the empty and doubly occupied states respectively. Here $|B\rangle$ and $|AB\rangle$ represent the bonding and anti-bonding states corresponding to the hybridization of the dots' orbitals. In this situation, it is convenient to introduce the following operators : $\sigma_{AB} = | AB\rangle\langle AB|$, $\sigma_B = |B\rangle\langle B| $, $\sigma_\emptyset = |\emptyset\rangle\langle \emptyset|$ and $\sigma_2 = |2\rangle\langle 2|$. We also have $\sigma_- = |B\rangle\langle AB| = \sigma_+^\dag$ and $\sigma_z = \sigma_{AB}-\sigma_B$. The Hamiltonian writes:

\begin{align} \label{EquationJaynesCummingsFull}
\mathcal{H} = \hbar\omega_c a^{\dag}a + \hbar\frac{\Omega}{2} \sigma_z + E_0  \sigma_0 + E_2 \sigma_2 + \hbar g \left( \sigma_- a^{\dag}+\sigma_+a \right)   \nonumber \\
+ \hbar \epsilon _{in} \left( e^{-i\omega_d t}a^{\dag } + e^{i\omega_d t}a \right) + \mathcal{H}_{Bath} + \mathcal{H}_{Bath}^{coupling} 
\end{align}

where $\epsilon_{in}$ is related to the microwave drive amplitude at the input of the cavity and $\omega_d$ its pulsation. The energy difference $\Omega=\sqrt(4 t^2+\epsilon^2)$ between bounding and anti-bounding states depends on the DQD hopping parameter $t$ and the inter-dot energy detuning $\epsilon$. Importantly, $\epsilon$ is controlled via DC gates voltages. The Hamiltonian $\mathcal{H}_{Bath}$ describes environmental degrees of freedom like electronic leads ($\mathcal{H}_{lead}=\sum_{q,r}\hbar\omega_{q,r} b_{q,r}^{\dag}b_{q,r}$ with $b_{q,r}$ the creation fermionic operator in the lead $r$) but also phonons, fluctuators, and the external photonic modes of the microwave cavity.  It controls the decoherence processes of the DQD-cavity system. The term $\mathcal{H}_{Bath}^{coupling}$ couples the baths to the system. One can write the coupled equations of motion for the charge qubit-cavity system in a transport situation, i.e. when a finite bias is applied to the source-drain electrodes of the double quantum dot. In the following, we only consider the coupling to the electronic bath and the dephasing term $\Gamma_\phi$ arising for example from low frequency charge noise acting on the detuning $\epsilon$, yielding a $\langle \sigma_z \rangle$ term. In the rotating frame of the driving field which oscillates at $\omega_d$, the system of equations to be solved within the rotating wave approximation is :

\begin{align}
& \frac{d}{dt} \langle a \rangle & = \ \ \ & - (\kappa/2+ i\Delta_{cd}) \langle a \rangle -i \epsilon_{in}-i g \langle\sigma_-\rangle & \label{EquationFullEOMa} \\
& \frac{d}{dt} \langle \sigma_- \rangle  & = \ \ \ & -(\gamma/2+\Gamma_\phi+i \Delta)\langle \sigma_- \rangle + i g \langle a (\sigma_{AB}-\sigma_B) \rangle  & \label{EquationFullEOMsigmaminus} \\
& \frac{d}{dt} \langle \sigma_{AB} \rangle  & = \ \ \ & - i g (\langle a \sigma_+\rangle-\langle a^\dagger \sigma_-\rangle)+\sum_{i\neq AB} \left( \Gamma_{AB \leftarrow i}\langle \sigma_{i} \rangle- \Gamma_{i \leftarrow AB}\langle \sigma_{AB} \rangle \right) & \label{EquationFullEOMsigmaAB} \\
& \frac{d}{dt} \langle \sigma_{B} \rangle  & = \ \ \ & i g (\langle a \sigma_+\rangle-\langle a^\dagger \sigma_-\rangle) + \sum_{i\neq B} \left( \Gamma_{B \leftarrow i}\langle \sigma_{i} \rangle - \Gamma_{i \leftarrow B}\langle \sigma_B \rangle \right) & \label{EquationFullEOMsigmaB} \\
& \frac{d}{dt} \langle \sigma_j \rangle &  = \ \ \ & \sum_{i\neq j} \left( \Gamma_{j \leftarrow i}\langle \sigma_i \rangle - \Gamma_{i \leftarrow j}\langle \sigma_j \rangle \right) & \label{EquationFullEOMsigmaj}
\end{align}

where $\Delta=\Omega-\omega_d$ is the qubit-drive detuning, $\Delta_{cd}=\omega_c-\omega_d$, $\kappa$ is the total decay rate of the cavity. The coupling to the reservoir continuum $\Gamma_{\alpha\leftarrow\beta}$ is determined by a Fermi's golden rule:

\begin{align}
& \Gamma_{\alpha\leftarrow\beta} = \Gamma_{\alpha\leftarrow\beta}^L + \Gamma_{\alpha\leftarrow\beta}^R \nonumber \\
& \Gamma_{\alpha\leftarrow\beta}^{r=1(2)} = \frac{2\pi}{\hbar} \lvert \gamma_r \rvert^2 \nu_r f_r \left( E_\alpha - E_\beta \right)   
\end{align}

where $\gamma_{r=1(2)}$ is the bare coupling rate to the reservoir $1(2)$, $\nu_{r=1(2)}$ is the density of state in reservoir $1(2)$ and $f_{r=1(2)}\left( E_\alpha - E_\beta \right) $ its Fermi function taken at energy difference between states $|\alpha\rangle$ and $|\beta\rangle$.

In order to obtain a closed set of equations, we make use of a semi-classical approximation for the cavity field which leads to: $\langle a (\sigma_{AB}-\sigma_B)\rangle \approx \langle a \rangle\times\langle(\sigma_{AB}-\sigma_B)\rangle$ and $\langle a \sigma_+\rangle \approx \langle a \rangle \times \langle \sigma_+\rangle$. This is justified in our case since we generally perform measurements with a number of photons in the cavity of the order of few $10-100$. In the stationary regime, equations \ref{EquationFullEOMsigmaminus} and \ref{EquationFullEOMa} yield:

\begin{align}
& \langle \sigma_- \rangle = \frac{\chi}{g} \langle a \rangle \langle \sigma_z \rangle \label{EquationSigmaminus} \\
& \langle a \rangle =  \frac{-i\epsilon_{in}}{i\Delta_{cd} + \frac{\kappa}{2} + i \chi \langle \sigma_z \rangle } \label{Equationa}
\end{align}

where $\chi$ is the charge susceptibility of the system:

\begin{equation} \label{EquationSusceptibility}
\chi = \frac{(g_0 \sin \theta)^2}{-i(\gamma/2+\Gamma_\phi)+ \Delta} = \frac{g^2}{-i \Gamma_2 + \Delta}
\end{equation}

where $\Gamma_2=\gamma/2+\Gamma_\phi$ is the inverse of the $T_2^*$ time of the charge qubit. Expression \ref{Equationa} yields a cavity frequency shift of the form

\begin{equation} \label{EquationDeltafc}
\Delta f_c =\Re e [\chi] \langle \sigma_z \rangle
\end{equation}

We recover here the traditional expression of the cavity frequency shift [6], but this is now valid in an electronic transport situation. The expression of  $\langle \sigma_z \rangle=\langle \sigma_{AB}-\sigma_B \rangle$ stems for the system of equation, arising from equations \ref{EquationFullEOMsigmaAB} to \ref{EquationFullEOMsigmaj}:

\begin{align} \label{Equationfulltheory}
-2 g^2 \Im m [\chi] \langle \sigma_{AB}-\sigma_B \rangle  \left\lvert \langle a \rangle \right\rvert^2  &= \sum_{i\neq AB} \left( \Gamma_{AB \leftarrow i}\langle \sigma_{i} \rangle- \Gamma_{i \leftarrow AB}\langle \sigma_{AB} \rangle \right) \nonumber & \\
2 g^2 \Im m [\chi] \langle \sigma_{AB}-\sigma_B \rangle \left\lvert \langle a \rangle \right\rvert^2 &= \sum_{i\neq B} \left( \Gamma_{B \leftarrow i}\langle \sigma_{i} \rangle - \Gamma_{i \leftarrow B}\langle \sigma_B \rangle \right)   \nonumber & \\
\langle \sigma_2 \rangle &= \frac{\Gamma_{2\leftarrow AB}\langle \sigma_{AB} \rangle+\Gamma_{2\leftarrow B}\langle \sigma_B \rangle}{\Gamma_{B\leftarrow 2}+\Gamma_{AB\leftarrow 2}} \nonumber & \\
\langle \sigma_\emptyset \rangle &= 1-\langle \sigma_2\rangle-\langle \sigma_B \rangle - \langle \sigma_{AB} \rangle  & 
\end{align}

The overall set of equations thus accounts for the effects arising from electronic transport and microwave drive of the charge qubit populations and allows the computation of the resulting frequency shift. Note that the semi-classical decoupling scheme described in this section can be refined in order to account for lasing, i.e. coherent photon-emission due to transitions between the AB and B states, which will be discussed experimentally in section \ref{Section:PhotonEmission}. Further details on these semiclassical calculation techniques can be found for instance in Refs [40,41].

    \section{Charge dynamics in hybrid double dot - resonator devices}

  		\subsection{Out of equilibrium transport and strong microwave drive}

In the linear regime, the hierarchy of the energy scales $eV_{SD} < k_BT \ll \Omega $ and $\Im m [\chi] n_{ph} \ll \gamma$ ensures that the system is on average close to its ground state, $\sigma_z \approx -1$. We can then determine the microwave phase response of the cavity at $f_c$, which directly reveals the DQD-induced cavity frequency shift:

\begin{equation} \label{EquationLinearRegime}
\Delta f_c = -\Re e [\chi] = -g^2 \frac{\Delta}{\Gamma_2^2 + \Delta^2}
\end{equation}

where $\Delta$ (or equivalently $\epsilon$) is directly controlled by DC gate voltages. Note that equation \ref{EquationLinearRegime} directly gives the behavior depicted in figure \ref{Figure:DQD}(C).

	\paragraph{Finite bias}
	
\begin{figure}\centering
\includegraphics[width=0.55\textwidth]{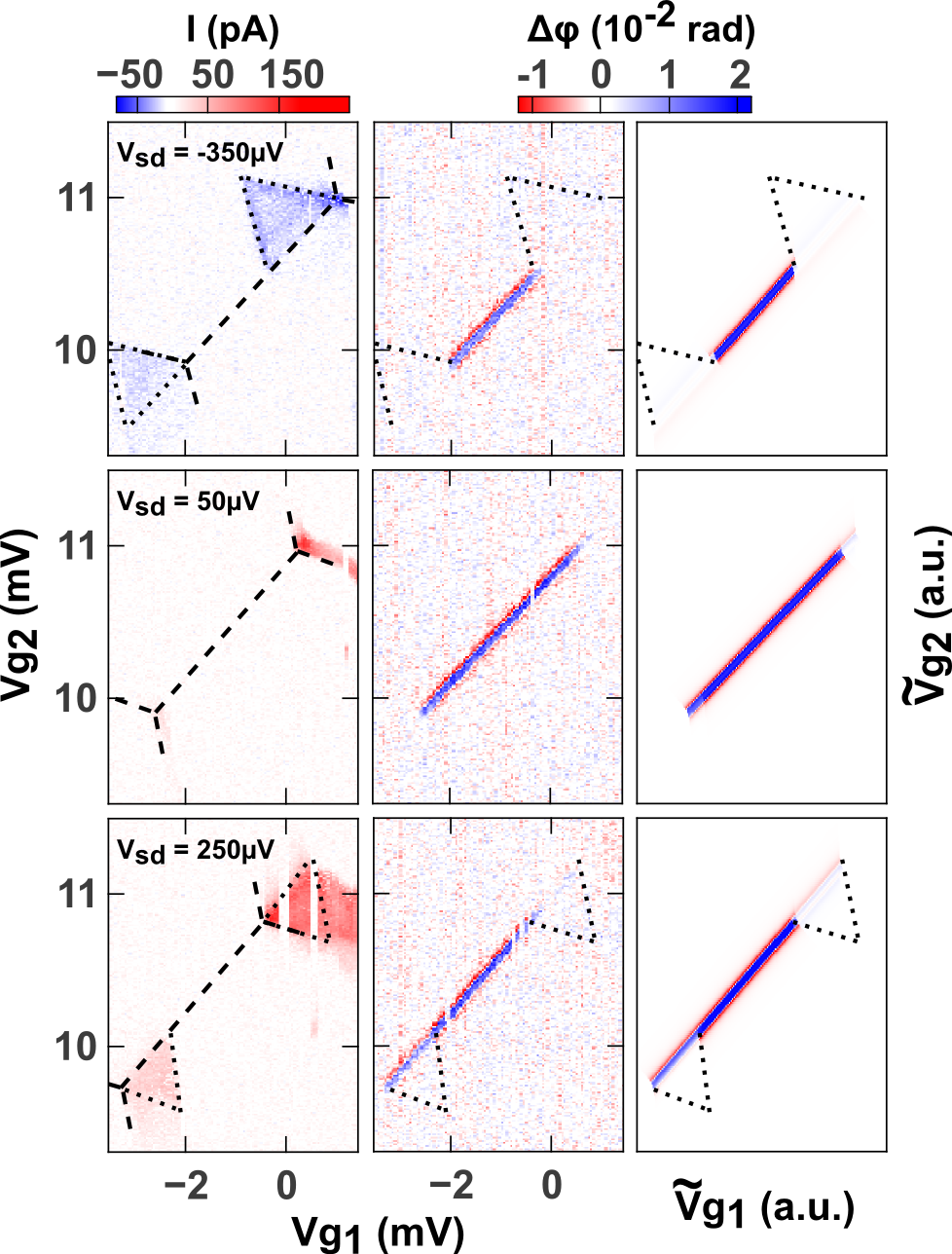}
\caption{Measured DC current through the DQD (first column), measured microwave phase (second column) and theory for microwave phase (third column) of the device at three different bias, as a function of the dots' gate voltages $V_{g1}$ and $V_{g2}$. Big dashed lines outline the charge stability diagram of the double quantum dot. The direct current signal shows the characteristic bias triangles (marked with small dashed lines) developing next to the triple points. The phase signal is unchanged between the bias triangles, where the charge remains blockaded, whereas it is modified in the regions where transport is allowed. Source: [37]}
\label{Figure:Charge:PRBFig3}
\end{figure}

When a finite bias is applied to the double quantum dot, electronic transport sets in and  $ \langle \sigma_\emptyset \rangle \neq 0 $ and $ \langle \sigma_2 \rangle  \neq 0$ in general. Knowing the charge susceptibility $\chi$ from measurements in the linear regime, we now have a direct measurement of the qubit z projection $\langle \sigma_z \rangle$ when a finite electronic current is driven through the double dot. Figure \ref{Figure:Charge:PRBFig3} shows such an experiment [37].

For $V_{SD} = 50 \mu V$, the phase contrast seems weakly affected despite non linear regime imposed by a bias larger than qubit splitting $\left( eV_{SD} > 2t \approx 40 \mu V \right)$. This is only because bias triangles appear to be small compared to the total distance between the triple points (determined by the mutual charging energy between the dots, $\approx 800\mu eV$ here). At $V_{SD} = 250 \mu V$, the length of the line between the triangle diminishes, and the phase contrast becomes weak under the top red triangle and a moderate under the bottom red triangle. This means that $\langle\sigma_z\rangle$ is strongly reduced under the top triangle (equal population for the bonding and anti-bonding states) whereas it stays finite (negative) under the bottom triangle. This difference of $\langle\sigma_z\rangle$ in the two triangles for positive bias reveals asymmetric dot-lead couplings. This shows the interest of the microwave phase signal in this out of equilibrium situation. For opposite bias ($V_{sd}=-350\mu V$), $\langle\sigma_z\rangle$ goes to zero under both triangles as illustrated in the top panel of figure \ref{Figure:Charge:PRBFig3}. This reveals symmetric and stronger dot-lead couplings of the DQD to both the leads at negative bias. As shown in the rightmost panels of figure \ref{Figure:Charge:PRBFig3}, we are able to reproduce the observed features with the theory which is developed in section \ref{SectionChargeEOM}. Reproducing these features strongly constraints the bare lead couplings at positive and negative biases. Although it is not as restrictive, the internal relaxation rate $\gamma$ is also constrained and this allows the differentiation of pure dephasing rate $\Gamma_\phi$ from decoherence rate $\Gamma_2^*=\gamma/2 + \Gamma_\phi$. In the device presented in Ref. [37], for the measured charge decoherence rate $\Gamma_2^*/2\pi=450 MHz$, the estimate of the relaxation $\gamma\simeq 300MHz$ yields $\Gamma_\phi \simeq 300MHz$.

	\paragraph{Finite microwave power} \label{SectionCavityPower}
	
The number of photons $n_{ph}$ in the cavity being  $|\langle a \rangle|^2$, we can compute the cavity readout power dependence in the case of no transport $ \langle \sigma_\emptyset \rangle = \langle \sigma_2 \rangle =0$ (we neglect thermal excitation of the AB state since $k_BT \ll \Omega$ and we neglect the self-consistency for determining the number of photons arising from the set (\ref{EquationFullEOMa},\ref{EquationFullEOMsigmaAB},\ref{EquationFullEOMsigmaB},\ref{EquationFullEOMsigmaj},\ref{EquationFullEOMsigmaminus}) due to the weak coupling strength $g_0$):

\begin{equation} \label{EquationPowersigmaZ}
\langle\sigma_z\rangle = \langle\sigma_{AB}-\sigma_B\rangle = \frac{-1}{1+4 \Im m [\chi] n_{ph} /\gamma}
\end{equation}

\begin{figure}\centering
\includegraphics[width=0.5\textwidth]{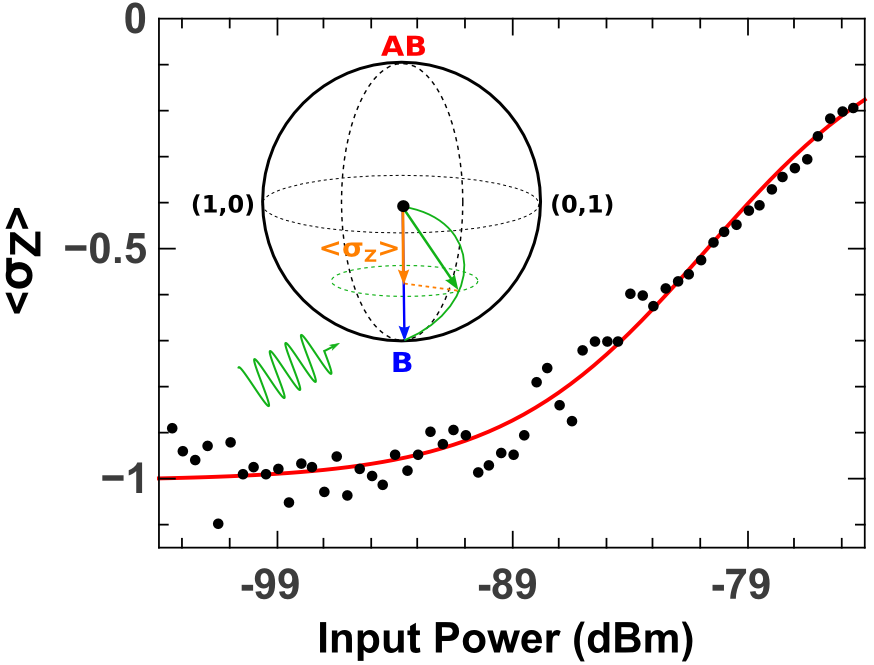}
\caption{Measured $\langle \sigma_z \rangle$ value (black points) obtained from the phase variation as a function of the estimated microwave power at the input of the cavity. Red line is theory described in section \ref{SectionCavityPower}. Inset : Bloch sphere of the charge qubit with bonding and anti-bonding sates. A large number of readout photons weakly detuned from the qubit excites transitions on average, and imposes $\langle \sigma_z \rangle > -1$. At the inflection point, Equation \ref{EquationPowersigmaZ} yields $\langle \sigma_z \rangle = -1/2$ when $4 \Im m [\chi] n_{ph}=\gamma$. Source: [37]}
\label{Figure:Charge:PRBFigPower}
\end{figure}

This formula yields the red solid line fitting the power dependence of $\langle \sigma_z \rangle$ in figure \ref{Figure:Charge:PRBFigPower}. The power dependence of the phase contrast at zero detuning allows one to determine the ratio between the relaxation rate $\gamma$ and the cavity photon number $n_{ph}$ at a given power. On average photons excite the qubit populations, thereby reducing the value of $\langle\sigma_z\rangle$. Since the photons \textit{drive} the effective spin, the efficiency of this process is directly related to the relaxation rate of the charge states. A direct measurement of the expectation value $\langle\sigma_z\rangle$ is displayed in figure \ref{Figure:Charge:PRBFigPower} (from [37]) at zero inter-dot detuning.

The average projection $\langle\sigma_z\rangle$ increases from its ground state value $-1$ up to $\approx -0.2$. Equation \ref{EquationPowersigmaZ} implies $4 \Im m [\chi] n_{ph} /\gamma=1$ when $\langle\sigma_z\rangle=-0.5$ and $\Im m [\chi]$ is determined from the low power study. We can in principle determine directly $\gamma$ provided $n_{ph}$ is accurately known, but precise knowledge of attenuation on such setup is in general non trivial. One can actually think of the opposite and use a precise knowledge of the qubit relaxation (i.e. through coherent manipulations) to determine $n_{ph}$ using equation \ref{EquationPowersigmaZ}.

		\subsection{Detection of spin blockaded states with a high Q, GHz resonator}

As demonstrated in the previous section, the DQD charge qubit populations are affected by electronic transport taking place in the dots, and this can be precisely measured via the phase and the resonator transmission. This particular fact has been be used to readout spin states [24] when the charge qubit populations depend on spin blockade effects [42,43]. Starting from a two-electrons charge configuration with one charge in both the dots $(1,1)$, and pulsing the gate voltages to drive the system in the $(0,2)$ configuration, the probability that the "left" electron tunnels on the "right" dot depends on whether the spin state of $(1,1)$ was a singlet or a triplet. Intuitively, if electrons spins are parallel, tunneling from $(1,1)$ to $(0,2)$ is spin blockaded; if they are anti-parallel, tunneling can take place. After the gate voltage pulse, the population of the singlet/triplet qubit is encoded in coherent superpositions of $(1,1)$ and $(0,2)$ and the charge occupation is thus spin dependent. In ref. [24], spin states are controlled with a classical AC field applied directly to the DQD gates via intrinsic spin-orbit interaction of a InAs nanowire. The resonator is thus used as a high frequency charge sensor to readout spin-orbit qubit states. This could for instance be used to readout the states of several spin-orbit qubits within a single resonator.

  		\subsection{Noise measurements}

The amount of charge noise in a solid state environment is very important for future development of devices exploiting degrees of freedom such as spin or valley. It is a priori the limiting factor for dephasing time of spin qubits in the absence of nuclear spins.  

Because charge noise strongly affects the decoherence rate of the charge qubits considered here, one can make the assumption that it is the dominating mechanism for dephasing. This allows the use of experimental estimates of dephasing times $\Gamma_{\phi}$ to give an upper bound of the typical charge noise in a device. We use a simple semi-classical model for dephasing with $1/f$ charge noise [44,45]. At zero detuning, the system is insensitive to charge noise at first order in the charge fluctuation. At second order:

\begin{equation} \label{EquationChargeNoise}
\Gamma_\phi \approx \frac{d^2\Omega}{d\epsilon^2} \langle \sigma_\epsilon \rangle^2 \approx \frac{\langle \sigma_\epsilon \rangle^2}{2t}
\end{equation}

In carbon nanotubes, we typically obtain $\langle\sigma_\epsilon\rangle=5 \mu eV$. With a typical charging energy of $10meV$ in the device, one can convert $\langle\sigma_\epsilon\rangle$ into a charge noise of  $ 5 \times 10^{-4} e/\sqrt{Hz}$ at $1Hz$. This allows us to give an upper bound for the charge noise of $ 5-15 \times 10^{-4} e/\sqrt{Hz}$ at $1Hz$, depending on gate settings.

\begin{figure}\centering
\includegraphics[width=0.65\textwidth]{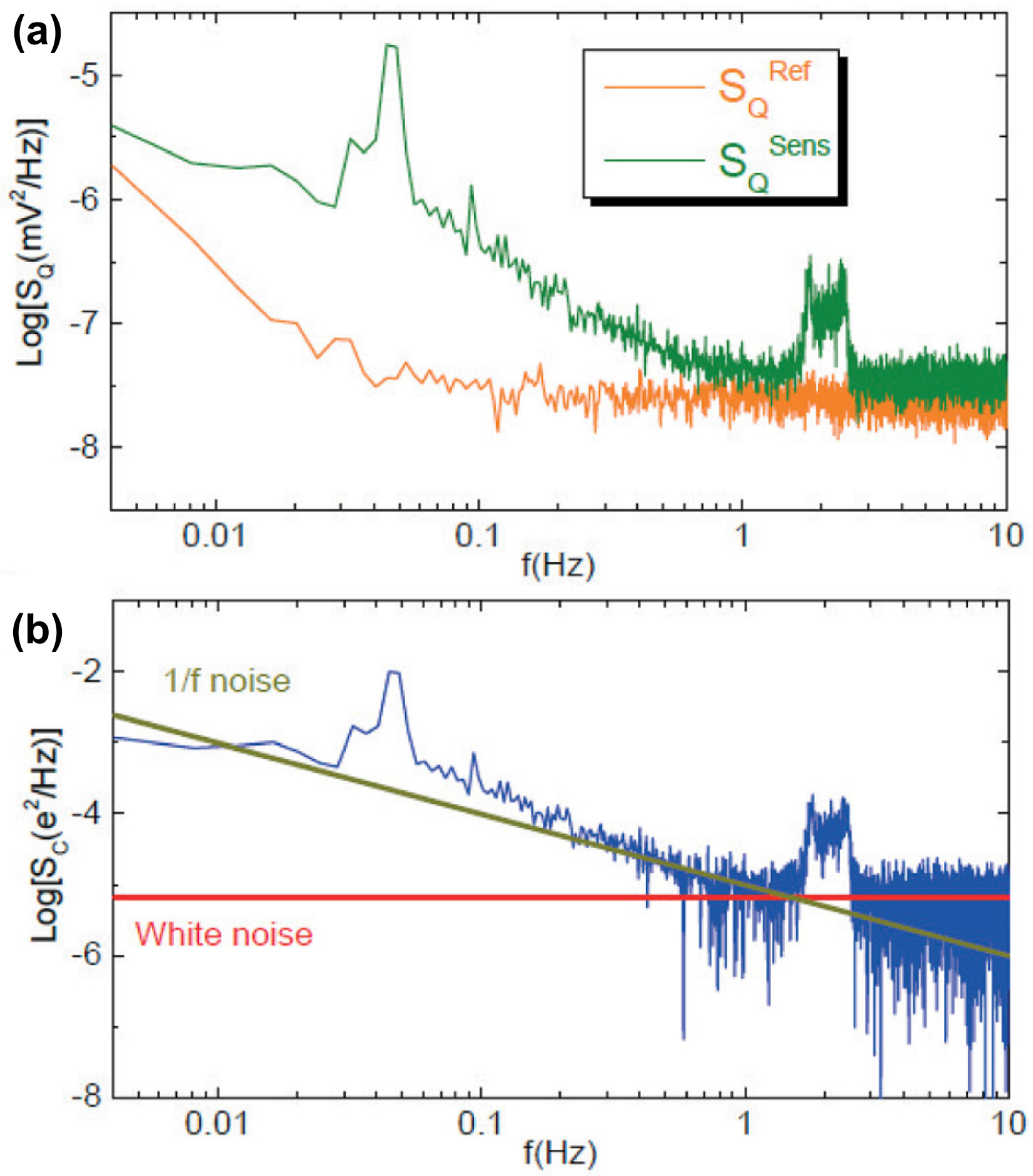}
\caption{\textbf{(a)} Quadrature noise spectrum $S_Q$ measured on a double quantum dot at two different inter-dot detuning corresponding to two different sensitivity to charge noise. \textbf{(b)} Charge noise spectral density $S_C$ of the surrounding environment of the double dot extracted from quadrature data sets. The two straight lines are guides to the eye corresponding to a white noise above $1 Hz$ or a $1/f$ divergence of the noise at low frequency. Source: [46]}
\label{Figure:Basset}
\end{figure}

One can also directly measure the noise power spectral density in the immediate environment of the double quantum dot. Charge noise, e.g. arising from fluctuators in the vicinity of the local gates, can be directly mapped onto the noise in the two quadratures of the cavity output field. Using a low noise (HEMT) amplifier, as shown in figure  \ref{Figure:Basset}, this allows one to obtain the full noise spectral density as a function of frequency and gate voltage [46]. For comparison, Ref. [46] has measured, in  GaAs-based DQD, an interdot detuning noise spectral density of $7.5 \mu eV/\sqrt{Hz} $ at $1Hz$, close to what has been found in carbon nanotubes, $5 \mu eV/\sqrt{Hz}$ in Ref. [37]. More recently, josephson parametric amplifiers have also been used to measure power spectral densities with a much greater efficiency [46,48].

		\subsection{Photon emission} \label{Section:PhotonEmission}

\begin{figure}\centering
\includegraphics[width=0.65\textwidth]{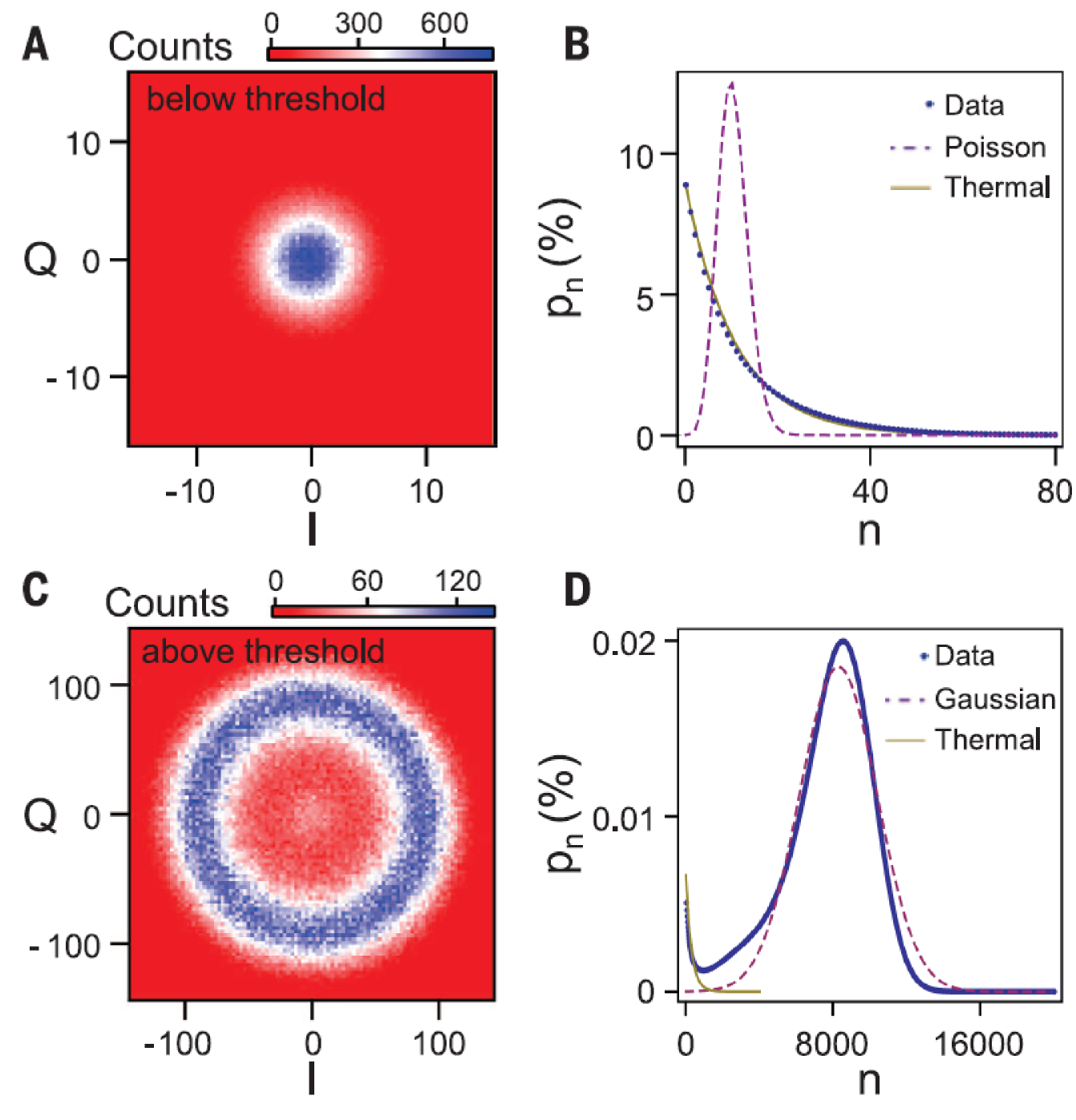}
\caption{(A) and (C) Histograms of the resonator output field quadratures acquired below  and above lasing threshold respectively. (B) and (D) Photon number distribution, extracted from the data in (A) and (C) respectively. Data is compared with thermal and Gaussian distributions in both cases, showing evidence of lasing above threshold. Source: [48]}
\label{Figure:Liu}
\end{figure}

Because charge doublets in double quantum dot are by nature non-linear systems (two level systems), they can be used to produce amplification when coupled to microwave light [47,48,49]. One can invert the charge doublet populations by using a DC bias voltage. Then, electrons can relax in the DQD by emitting a photon in the resonator (and also phonons in their environment). This has been observed first in DQD made out of InAs [47], and studied in more details in GaAs more recently [50]. Using two InAs-based double quantum dots in a resonator, lasing (or masing) has also recently been demonstrated [48] (see figure \ref{Figure:Liu}). The laser action was verified  by studying the statistics of the two quadratures of the emitted microwave field. Below the lasing threshold, this cavity field tomography reveals a thermal occupation of the photon states (figure \ref{Figure:Liu}.A). Above the threashold, the coherence of the photonic emission was observed (figure \ref{Figure:Liu}.C). Because these inter-dot transitions are widely tunable, they could be use to generate lasing in wide range of frequency, up to $THz$ [48].

	\section{Spin-photon coupling}

\begin{figure}\centering
\includegraphics[width=0.85\textwidth]{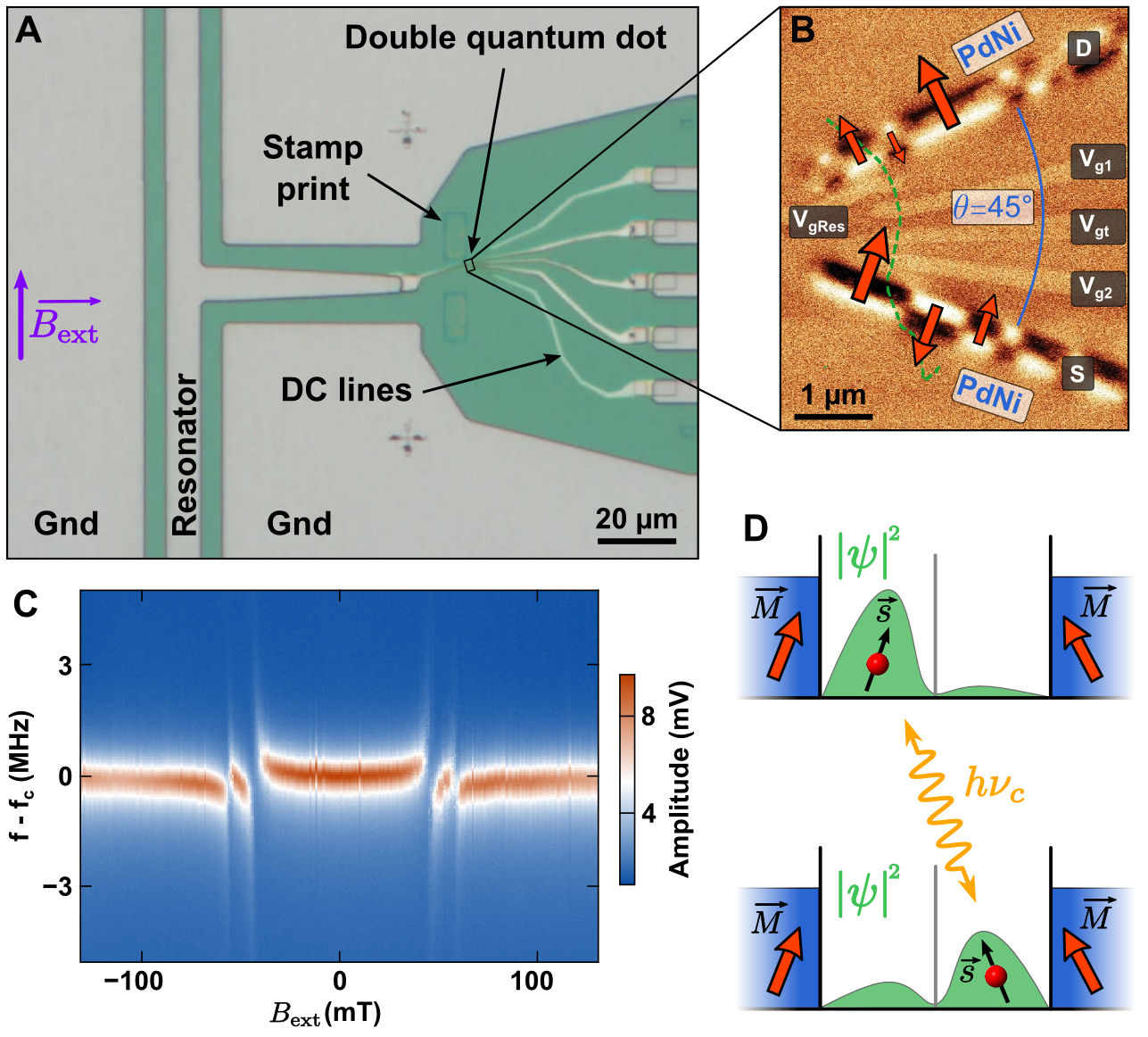}
\caption{(A) Optical micrograph showing the superconducting coplanar waveguide resonator and the double quantum dot device. (B) Magnetic force micrograph of the double quantum dot showing four (non-magnetic) top gates and the source S and drain D electrodes made out of a ferromagnetic alloy (PdNi). Black and white colors correspond to north and south poles of ferromagnetic domains. We indicate with a green dashed line the position of the carbon nanotube as it appears on the atomic force micrograph (not shown). (C) Microwave resonator spectrum as a function of external magnetic field. Two spin transitions, strongly dispersing with magnetic field, become resonant and hybridize with the resonator mode. (D) General principle of the coupling mechanism. The proximity of the noncollinear ferromagnets induces a different equilibrium spin orientation if an electron is localized in the left or in the right dot. Photons are coupled to transitions changing the localization of the wave function, hence coupled to transitions changing the spin orientation. Source: [8]}
\label{Figure:Science}
\end{figure}

In a solid-state context, the electronic spin degree of freedom is naturally much more isolated from its environment than the electronic charge. In nuclear spin free host materials, electron spins have long dephasing times. For instance $T_2^* > 100 \mu s$ has been observed for individual ${}^{31}P$ donor electron spin in a silicon based architecture [51], and $T_2^* \simeq 360 ns$ has been measured for a singlet-triplet qubit in Si/SiGe quantum dots [52]. The counter part of this weak coupling to the environment is the equivalently weak magnetic coupling of a single electronic spin to the electromagnetic field of a cavity, which is about $g_{spin}\simeq 50Hz$ in standard planar geometries [53,54]. In order to reach the strong coupling limit ($g_{spin}>2/T_2^*$ and $g_{spin}>2/ \kappa$), one must find a trick to increase $g_{spin}$, utilizing the large charge-photon coupling to create an effective spin electric dipole moment. To do so, proposals rely either on intrinsic properties such as Overhauser felds [12,14], natural spin-orbit coupling [11,24] or on extrinsically engineered coupling such as spin-charge entanglement using a Raman transition [13] or artificial spin-orbit coupling [10]. Ref. [10] proposes to use local effective Zeeman fields induced in a DQD by the interface with ferromagnetic reservoirs. This has been recently demonstrated experimentally using a carbon nanotube DQD (figure \ref{Figure:Science}), with a spin-photon coupling in the $MHz$ range [8]. In this experiment, $T_2^*$ is estimated to be $60ns$, which sets the system at the strong coupling threshold. As depicted in \ref{Figure:Science}D, spin-photon coupling is enabled by the hybridization of charge and spin states in the DQD, a key ingredient being the non-colinearity of the local effective fields in the two dots. This gives a cooperativity which is large enough to observe the hybridization of the resonator with DQD transitions which correspond dominantly to a spin reversal, see fig. \ref{Figure:Science}C. A detailed data analysis reveals that the measured coherence time is limited by charge noise, and could possibly be further improved by optimizing the spin-charge hybridization [10]. Furthermore, because the coupling to the cavity is gate tunable, pulsed measurement could access the pure spin coherence time in carbon nanotube, which has not been measured to date. 

    \section{Conclusion and perspectives}

The experiments described in this short review open up a wide spectrum of new studies in hybrid circuit QED with mesoscopic circuits such as quantum dots. Further progresses in the developpement of the spin/photon coupling could enable the distant coupling of spins through cavity photons, and various types of quantum spin manipulations. Furthermore, the versatility of nanofabrication techniques should enable to study a vary large variety of structures and situations. For a quantum dot with a single normal metal contact, a cavity would provide an accurate way to study the problem of the  universality of charge relaxation [28], which is expected at low temperatures for a purely capacitive dot/cavity coupling [4]. Nonlocal transport effects are also predicted, where for instance the finite bias applied to a given quantum dot should trigger electron tunneling in another distant quantum dot [55,56,57]. One could also use the tools of circuit QED to probe electronic entanglement generated by the splitting of Cooper pairs in double quantum dot setups [58,59], or even the peculiar properties of emergent Majorana fermions in superconducting/nanowire heterostructures [60,61]. Quantum computing schemes based on Majorana bound states coupled to microwave cavities have also been proposed recently [62]. Finally, it could be particularly interesting to extend further the ideas presented above to the THz spectral domain which would match the characteristic energy scales of interacting wires (in the Luttinger liquid regime) or quantum dots (in the Kondo regime).

\section*{References} 
[1] G. Binnig, C. F. Quate, Phys. Rev. Lett. 56, 930 (1986)

[2] R. J. Abraham, J. Fisher, P. Loftus, Introduction to NMR Spectroscopy (1992)

[3] D. J. Reilly, C. M. Marcus, M. P. Hanson, A. C. Gossard, Applied Physics Letter 91, 162101 (2007)

[4] A. Pretre, H. Thomas, and M. Buttiker, Phys. Rev. B 54, 8130 (1996)

[5] J. Gabelli, G. Feve, J.-M. Berroir, B. Placais, A. Cavanna, B. Etienne, Y. Jin, D. C. Glattli, Science 313, 499-502 (2006)

[6] A. Blais, R.-S. Huang, A. Wallraff, S. M. Girvin, and R. J. Schoelkopf, Phys. Rev. A 69, 062320 (2004)

[7] A. Wallraff, D. Schuster, A. Blais, L. Frunzio, R.-S. Huang, J. Majer, S Kumar, S. M. Girvin, and R. J. Schoelkopf, Nature 431, 162-167 (2004)

[8] J. J. Viennot, M. C. Dartiailh, A. Cottet and T. Kontos, Science 349, 408-411 (2015) 

[9] D. Loss, D. P. DiVincenzo, Phys. Rev. A 57, 120 (1998)

[10] A. Cottet, T. Kontos, Phys. Rev. Lett. 105, 160502 (2010)

[11] M. Trif, V. N. Golovach, and D. Loss, Phys. Rev. B 77, 045434 (2008)

[12] P.-Q. Jin, M. Marthaler, A. Shnirman, and G. Schon, Phys. Rev. Lett. 108, 190506 (2012)

[13] L. Childress, A. Sorensen, and M. Lukin, Phys. Rev. A 69, 042302 (2004)

[14] G. Burkard and A. Imamoglu, Phys. Rev. B 74, 041307(R) (2006)

[15] S. Hermelin, S. Takada, M. Yamamoto, S. Tarucha, A. D. Wieck, L. Saminadayar, C. Bauerle, and T. Meunier, Nature 477, 435?438 (2011)

[16] L. Trifunovic, O. Dial, M. Trif, J. R. Wootton, R. Abebe, A. Yacoby, and D. Loss, Phys. Rev. X 2, 011006 (2012)

[17] Y. Kubo, F. Ong, P. Bertet, D. Vion, V. Jacques, D. Zheng, A. Dreau, J.-F. Roch, A. Auffeves, F. Jelezko, J. Wrachtrup, M. Barthe, P. Bergonzo, and D. Esteve, Phys. Rev. Lett. 105, 140502 (2010)

[18] D. Schuster, A. Sears, E. Ginossar, L. DiCarlo, L. Frunzio, J. Morton, H. Wu, G. Briggs, B. Buckley, D. Awschalom, and R. J. Schoelkopf, Phys. Rev. Lett. 105, 140501 (2010)

[19] Y. Tabuchi, S. Ishino, T. Ishikawa, R. Yamazaki, K. Usami, Y. Nakamura, Phys. Rev. Lett. 113, 083603 (2014)

[20] A D O'Connell, M. Hofheinz, M. Ansmann, R. C. Bialczak, M. Lenander, E. Lucero, M. Neeley, D. Sank, H. Wang, M. Weides, J. Wenner, J. M. Martinis, and A. N. Cleland, Nature 464, 697-703 (2010)

[21] D Teufel, T Donner, D. Li, J. W. Harlow, M .S. Allman, K. Cicak, A. J. Sirois, J. D. Whittaker, K. W. Lehnert, and R. W. Simmonds, Nature 475, 359-363 (2011)

[22] J. Bochmann, A. Vainsencher, D. Awschalom, and A. N. Cleland, Nat. Phys. 9, 712-716 (2013)

[23] T. Frey, P. J. Leek, M. Beck, K. Ensslin, A. Wallraff, T. Ihn, Appl. Phys. Lett. 98, 262105 (2011)

[24] K. D. Petersson, L. W. McFaul, M. D. Schroer, M. Jung, J. M. Taylor, A. A. Houck, and J. R. Petta, Nature 490, 380-383 (2012)

[25] J. J. Viennot, J. Palomo, T. Kontos, Appl. Phys. Lett. 104, 113108 (2014)

[26] M.-L. Zhang, D. Wei, G.-W. Deng, S.-X. Li, H.-O. Li, G. Cao, T. Tu, M. Xiao, G.-C. Guo, H.-W. Jiang, G.-P. Guo, Appl. Phys. Lett. 105, 073510 (2014)

[27] M. Goppl, A Fragner, M Baur, R. Bianchetti, S Filipp, J. M. Fink, P. J. Leek, G Puebla, L Steffen, and A. Wallraff, J. Appl. Phys. 104, 113904 (2009)

[28] A. Cottet, T. Kontos, B. Doucot, Phys. Rev. B 91, 205417 (2015)

[29] A. Cottet, C. Mora, T. Kontos, Phys. Rev. B 83, 121311(R) (2011)

[30] M.R. Delbecq, V. Schmitt, F. Parmentier, N. Roch, J. J. Viennot, G. Feve, B. Huard, C. Mora, A. Cottet, and T. Kontos, Phys. Rev. Lett. 107, 256804 (2011)

[31] T. Frey, P. J. Leek, M. Beck, J. Faist, A. Wallraff, K. Ensslin, T. Ihn, and M. Buttiker, Phys. Rev. B 86, 115303 (2012)

[32] L.E. Bruhat, J.J. Viennot, M.C. Dartiailh, M.M. Desjardins, T. Kontos, A. Cottet, Phys. Rev. X 6 (2016) 021014.

[33] M.R. Delbecq, L. E. Bruhat, J. J. Viennot, S Datta, A. Cottet, and T Kontos, Nat. Comm. 4, 1400 (2013)

[34] W. van der Wiel, S. De Franceschi, J. M. Elzerman, T. Fujisawa, S. Tarucha, and L. P. Kouwenhoven, Rev. Mod. Phys. 75, 1 (2002)

[35] T. Hayashi, T. Fujisawa, H. Cheong, Y. Jeong, and Y. Hirayama, Phys. Rev. Lett. 91, 226804 (2003)

[36] T. Frey, P. J. Leek, M. Beck, A. Blais, T. Ihn, K. Ensslin, and A. Wallraff, Phys. Rev. Lett. 108, 046807 (2012)

[37] J. J. Viennot, M.R. Delbecq, M. C. Dartiailh, A. Cottet, and T. Kontos, Phys. Rev. B 89, 165404 (2014)

[38] H. Toida, T. Nakajima, and S. Komiyama, Phys. Rev. Lett. 110, 066802 (2013). A. Wallraff, A. Stockklauser, T. Ihn, J. R. Petta, and A. Blais, Phys. Rev. Lett. 111, 249701 (2013)

[39] M. D. Schroer, M. Jung, K. D. Petersson, and J. R. Petta, Phys. Rev. Lett. 109, 166804 (2012)

[40] S. Andre, V. Brosco, M. Marthaler, A.Shnirman, G. Schon, Phys. Scr. T 137, 014016 (2009)

[41] M. Kulkarni, O. Cotlet, and H. E. Tureci, Phys. Rev. B 90, 125402 (2014)

[42] K Ono, D G Austing, Y Tokura, and S. Tarucha, Science 297, 1313-1317 (2002)

[43] R. Hanson, J. R. Petta, S. Tarucha, and L. M. K. Vandersypen, Rev. Mod. Phys. 79, 1217 (2007)

[44] A. Cottet, PhD thesis (2002)

[45] G. Ithier, E. Collin, P. Joyez, P. Meeson, D. Vion, D. Esteve, F. Chiarello, A. Shnirman, Y. Makhlin, J. Schriefl, and G. Schon, Phys. Rev. B 72, 134519 (2005)

[46] J. Basset, A. Stockklauser, D.-D. Jarausch, T. Frey, C. Reichl, W. Wegscheider, A. Wallraff, K. Ensslin, T. Ihn, Appl. Phys. Lett. 105, 063105 (2014)

[47] Y.-Y. Liu, K. D. Petersson, J. Stehlik, J. M. Taylor, and J. R. Petta, Phys. Rev. Lett. 113, 036801 (2014)

[48] Y.- Y. Liu, J. Stehlik, C. Eichler, M. J. Gullans, J. M. Taylor, J. R. Petta, Science 347,  285-287 (2015)

[49] P.-Q. Jin, M. Marthaler, J. H. Cole, A. Shnirman, and G. Schon, Phys. Rev. B 84, 035322 (2011).

[50] A. Stockklauser, V.?F. Maisi, J. Basset, K. Cujia, C. Reichl, W. Wegscheider, T. Ihn, A. Wallraff, and K. Ensslin, Phys. Rev. Lett. 115, 046802 (2015)

[51] J. T. Muhonen, J. P. Dehollain,	A. Laucht, F. E. Hudson, R. Kalra, T. Sekiguchi, K. M. Itoh, D. N. Jamieson, J. C. McCallum, A. S. Dzurak and A. Morello, Nature Nano. 9, 986-991 (2014)

[52]  B. M. Maune, M. G. Borselli, B. Huang, T. D. Ladd, P. W. Deelman, K. S. Holabird,	A. A. Kiselev,	I. Alvarado-Rodriguez,	R. S. Ross,	A. E. Schmitz,	M. Sokolich, C. A. Watson, M. F. Gyure and A. T. Hunter, Nature 481, 344-347 (2012)

[53] A. Bienfait, J. J. Pla, Y. Kubo, M. Stern, X. Zhou, C. C. Lo, C. D. Weis, T. Schenkel, M.L.W. Thewalt, D. Vion, D. Esteve, B. Julsgaard, K. Moelmer, J.J.L. Morton and P. Bertet, arXiv:1507.06831 (2015)

[54] A. Bienfait, J.J. Pla, Y. Kubo, X. Zhou, M. Stern, C.C. Lo, C.D. Weis, T. Schenkel, D. Vion, D. Esteve, J.J.L. Morton and P. Bertet, arXiv:1508.06148 (2015)

[55] C. Bergenfeldt, and P. Samuelsson, Phys. Rev. B 87, 195427 (2013)

[56] N. Lambert, C. Flindt, and F. Nori, Europhys. Lett. 103, 17005 (2013)

[57] L. D. Contreras-Pulido, C. Emary, T. Brandes, and R. Aguado, New J. Phys. 15, 095008 (2013)

[58] A. Cottet, T. Kontos, A. L. Yeyati, Phys. Rev. Lett. 108, 166803 (2012)

[59] A. Cottet, Phys. Rev. B 90, 125139 (2014) 

[60] A. Cottet, T. Kontos, B. Doucot, Phys. Rev. B 88, 195415 (2013) 

[61] M. Trif and Y. Tserkovnyak, Phys. Rev. Lett. 109, 257002 (2012).

[62] F. Hassler, A. R. Akhmerov and C. W. J. Beenakker, New J. Phys. 13, 095004 (2011).

\end{document}